\documentclass[preprint,11pt]{JHEP3}

\JHEPspecialurl{http://jhep.sissa.it/JOURNAL/JHEP3.tar.gz}

\usepackage{epsfig,multicol,amsmath}

\usepackage{bm}  
\usepackage{amsmath}     
\usepackage{epsfig,epsf}
\usepackage{rotating}
\usepackage{cite}

\def\beq{\begin{equation}}
\def\eeq{\end{equation}}
\def\bea{\begin{eqnarray}}
\def\eea{\end{eqnarray}}

\def\eq#1{{Eq.~(\ref{#1})}}
\def\fig#1{{Fig.~\ref{#1}}}
\newcommand{\bas}{\bar{\alpha}_S}
\newcommand{\as}{\alpha_S}

\newcommand{\Lb}{\left(}
\newcommand{\Rb}{\right)}
\newcommand{\h}{\frac{1}{2}}
\parskip=\itemsep               

\newcommand{\nn}{\nonumber}

\def\pom{{I\!\!P}}

\begin{boldmath}
\title{Dipole-dipole scattering in  CGC/saturation approach at high energy: summing Pomeron loops}
\end{boldmath}
\author{\Large 
Eugene\, Levin${}^{a, b}$ \thanks{Email: leving@post.tau.ac.il, eugeny.levin@usm.cl.}\\
${}^a$ \, Department of Particle Physics, School of Physics and Astronomy,\\
Tel Aviv University, Tel Aviv, 69978, Israel\\
${}^b$\, Departamento de F\'\i sica,
Universidad T$\acute{e}$cnica Federico Santa Mar\'\i a   and\\
Centro Cient\'\i fico-Tecnol$\acute{o}$gico de Valpara\'\i so,
Casilla 110-V,  Valpara\'\i so, Chile\\
}

\abstract
{In this paper we demonstrate that the dense system of partons (gluons) can be produced in dilute-dilute system scattering, using the example  of dipole-dipole collisions. This increase in density stems from the intensive gluon cascades that can be described by the enhanced BFKL Pomeron diagrams (Pomeron loops).
For the first time we found the analytical solution to the equation for diffraction production in the dipole-dense parton system scattering,  using the simplified BFKL kernel.  Having this solution as well as the solution to Balitsky- Kovchegov equation we developed technique that allowed us to calculate the total cross section, cross sections for single and double diffractions in the MPSI approximation. Calculating inclusive production and two gluon correlations we see that the dense and strongly correlated system of gluons can be produced at high energy in the dipole-dipole scattering. }

\dedicated{PACS: 12.38-t, 12.38.Cy,1 2.38.Lg, 13.60.Hd, 24.85.+p, 25.30.Hm}
\preprint{TAUP 2971/13 \\
{\tt }\\
\today}

\begin{document}
\section{ Introduction}
LHC data support the assumption that the dense system of partons is produced in the proton-proton collisions at high energy. 
Such dense system of partons naturally appears in the CGC/saturation approach to high energy QCD\cite{GLR,MUQI,MV,BK,JIMWLK,REV}. The success in description of both the general properties of the bias event \cite{LERE} and the long range rapidity angular correlations in the framework of CGC/saturation approach \cite{COR} makes this assumption a working hypothesis 
which allows us to look at hadron-hadron, hadron-nucleus and nucleus-nucleus interactions from the unique point of view. We wish to single out two sets of the experimental data which confirm the hypothesis. The first one is the measurement of the long range rapidity correlations in the azimuthal angle between two produced hadrons \cite{CMSCOR} which have the same pattern as such correlations in proton-nucleus\cite{RDGPA} and nucleus-nucleus collisions\cite{RDGAA}. The second set of the data is the measurement of the double parton interaction (DPI)\cite{DPI}. In these experiments the double inclusive cross sections of two pair of back-to-back jets with momenta $p_{T,1}$ and $p_{T,2}$,  were measured with rapidities  of two pairs ($y_1$ and $y_2$)  which are close to each other ($y_1 \approx y_2$). These pairs can be produced only from two different parton showers.  The data were parameterized in the form
\beq \label{XSEFF}
\frac{d \sigma}{d y_1 d^2 p_{T,1} d y_2 d^2 p_{T,2}} \,\,=\,\,\frac{m}{2 \sigma_{eff}}\,\frac{d \sigma}{d y_1 d^2 p_{T,1}}\frac{d \sigma}{d y_2 d^2 p_{T,2}} 
\eeq
where $m =2 $ for different pairs of jet and $m=1$ for identical pairs.  One can calculate the rapidity correlation function using \eq{XSEFF}
\beq \label{R}
R\Lb y_1, y_2\Rb\,\,=\,\,\frac{\frac{1}{\sigma_{in}}\frac{d \sigma}{d y_1 d^2 p_{T,1} d y_2 d^2 p_{T,2}}}{\frac{1}{\sigma_{in}}\frac{d \sigma}{d y_1 d^2 p_{T,1}}\,\frac{1}{\sigma_{in}}\frac{d \sigma}{d y_2 d^2 p_{T,2}} }\,\,-\,\,1\,=\frac{\sigma_{in}}{\sigma_{eff}}\,\,-\,\,1\,\,\approx\,\,2
\eeq
For the above the estimates we use $\sigma_{eff} $=\,12 - 15\,mb (see Refs. \cite{DPI}) and $\sigma_{in} = \sigma_{tot} - \sigma_{el} - \sigma_{sd} - \sigma_{dd} \,\approx $\,50 mb for the energy $W = 7 \,TeV$ (see Ref.\cite{GLM} and references therein). Using 
that $y_1 \approx y_2 \approx 4\div 5 $ in ATLAS experiment at $W= 7\, TeV$ (see Ref.\cite{DPI}) and the estimates that one gluon jet decays in two hadrons \cite{LERE} we can evaluate the density of parton in rapidity , namely, $d N_{parton}/d y \approx 1$ from the inclusive cross sections measured at the LHC \cite{INCEXP}. Therefore, we can conclude that at $W= 7\, TeV$ the dense system of parton produced and these parton strongly interact with each other.

\begin{figure}
  \leavevmode
  \begin{tabular}{c }
      \includegraphics[width=14cm]{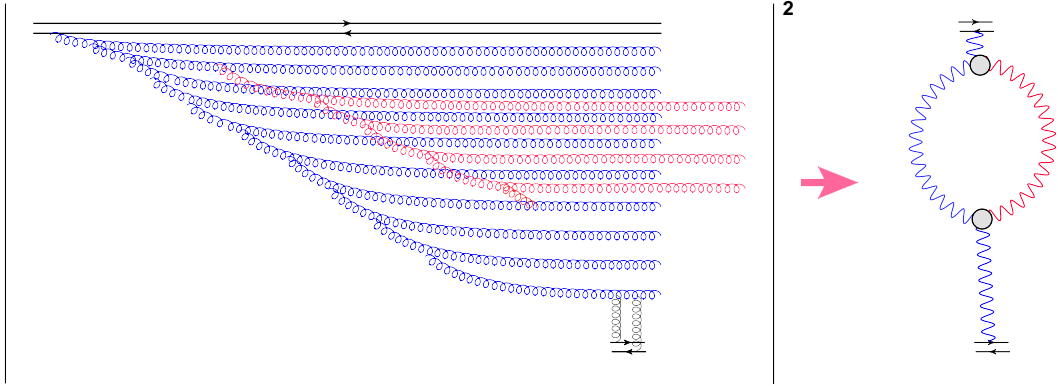}\\
   \end{tabular}
      \caption{The first enhanced BFKL Pomeron diagram for dipole-dipole scattering that leads to an  increase of the parton density. The helix lines show gluons while the solid lines are used for quarks and antiquarks. The picture is drawn in the l.r.f. where one of the dipoles is at the rest.}
      \label{endi}
\end{figure}

\begin{figure}
  \leavevmode
  \begin{tabular}{c }
      \includegraphics[width=14cm]{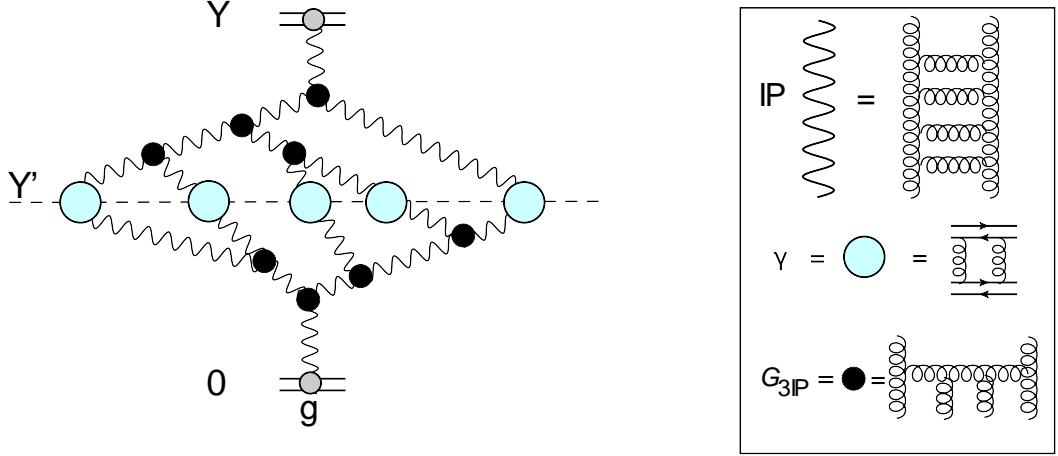}\\
   \end{tabular}
      \caption{ MPSI approximation:  all notations are shown in the insertion. Wavy lines denote the BFKL Pomerons, 
      the blob stands for the scattering amplitude of two dipoles with sizes:$ r_1$ and $r_2$ in the Born approximation (due to exchange of two gluons). $G_{3 \pom}$ is the triple BFKL Pomerons vertex. }
      \label{mpsi}
\end{figure}

 Since at low energy the proton consists of a moderate number of partons and can be considered as a dilute parton system, the only way, how the proton could become a dense system of partons, is due to intensive decay of partons inside the parton cascade (see \fig{endi}). In other words, we need to sum the enhanced BFKL Pomeron \cite{BFKL,RevLI} diagrams to create the dense system of partons. The first such diagram is shown in \fig{endi}.
Therefore, we face the difficult problem of summing enhanced diagrams in the framework of the BFKL Pomeron calculus \cite{BRAUN}.   The goal of this paper is to sum all enhanced diagrams for dipole-dipole scattering at high energy.
For solution of this problem we are going to exploit  Mueller-Patel-Salam-Iancu (MPSI) approximation\cite{MPSI} shown in \fig{mpsi}, following the procedure suggested in Ref.\cite{LMP}. One can see that the MPSI approximation is based on two key features of high density QCD: on $t$-channel unitarity and on the simple dipole cascade generated by dilute system (one upper or lower dipole in our case).
This cascade can be described by the simple generating functional \cite{MUCD,LELU} which is equivalent to 
Balitsky-Kovchegov\cite{BK} and JIMWLK(KLWMIG)\cite{JIMWLK} evolutions. In the BFKL Pomeron calculus this cascade corresponds to summation of fan Pomeron diagrams in the region of $Y - Y'\,\, \gg \,\,1$ and $Y' - 0 \,\,\gg\,\,1$ in \fig{mpsi}). Rapidity $Y'$ is artificial rapidity which does not enter the final answer. Indeed, the BFKL Pomeron has the following property from the $t$-channel unitarity \cite{GLR,MUSH} at any value of $Y'$:
\bea \label{POMTCH}
&&N_\pom\,=\,\int d^2 r_1 d^2 r_2 d^2 b_1 d^2 b_2 \,G_\pom\Lb r, r_1, \vec{b} - \vec{b}_1| Y - Y' \Rb \,\gamma\Lb r_1,r_2,\vec{b}_1 - \vec{b}_2\Rb \,\,G_\pom\Lb r_2, R, \vec{b}_2 |  Y' \Rb\,\,\nn\\
&&~~~~~~~~~~~~~~~~~~~~~~~~~~~~~~~~~~~~~~~~~~~~~~~~~~~~~~~~~~~~~~~~~~~~~~~~~~=\,\,\Lb \frac{\as^2}{4 \pi}\Rb^2 \,\,G_\pom\Lb r, R, \vec{b}| Y - 0 \Rb
\eea
In \eq{POMTCH} $N_\pom$ describes the dipole-dipoles scattering amplitude due to the single BFKL Pomeron exchange, $G_\pom$ denotes the Green function of the BFKL Pomeron. $\gamma\Lb r_1,r_2,\vec{b}_1 - \vec{b}_2\Rb$ is the amplitude  of  interaction of two dipoles with sizes: $r_1$ and $r_2$ at the impact parameter
$\vec{b}_1 - \vec{b}_2$ in the Born Approximation of perturbative QCD in which two dipoles interact due to exchange of two gluons.
$r$  and $R$ are  the sizes of the scattering dipoles.

In the next section we will find the amplitude for dipole-dipole scattering (or the resulting Green function of the BFKL Pomeron) at high energy using the approach proposed in Ref.\cite{LETU}.  
.

\section{Parton cascade of the  fast dipole}

Most of the material of this  section  is not new and has already appeared in \cite{LETU,KLT}. We include it here for completeness in order to present a coherent picture of the approach.

\subsection{Simplified BFKL kernel}

 BFKL kernel  is rather complicated and the analytical solution of the non-linear equation  with this kernel has not been found. In Ref.\cite{LETU} it was suggested to simplify the kernel by taking into account only log contributions. From formal point of view this simplification means that we consider only leading twist contribution to the BFKL kernel. Note that the full BFKL kernel includes all twists contributions.Actually we have two kinds of logs: $ \Big(\bas \ln\Lb r^2\,\Lambda^2_{QCD}\Rb\Big)^n$ outside of the saturation region ($r^2\,Q^2_s\Lb Y,b\Rb\,\,\equiv\,\,\tau\,\,\ll\,\,1$); and $  \Big(\bas \ln\Lb r^2\,Q^2_s\Lb Y,b \Rb \Rb\Big)^n$ inside the saturation domain ($\tau\,\gg\,1$). To sum all logs for $\tau \ll 1$ we can 
 simplify the BFKL kernel  $K\Lb r;r'\Rb$  in the following way\cite{LETU}, since $r'
 \gg r$ and $|\vec{r} - \vec{r}'| > r$
 \beq \label{K1}
 \int d^2 r' \,K\Lb r, r'\Rb\,\,\equiv\,\,\int d^2 r'  \frac{r^2}{r'^2\,\Lb \vec{r} - \vec{r}^{\,'}\Rb^2}\,\,\rightarrow\,\pi\, r^2\,\int^{\frac{1}{\Lambda^2_{QCD}}}_{r^2} \frac{ d r'^2}{r'^4}
\eeq

Inside of the saturation region where $\tau\,\,>\,\,1$ the logs are   originated from the decay of the large size dipole into one small size dipole  and one large size dipole.  However, the size of the small dipole is still larger than $1/Q_s$. This observation can be translated in the following form of the kernel
\beq \label{K2}
 \int d^2 r' \,K\Lb r, r'\Rb\,\,\rightarrow\,\pi\, \int^{r^2}_{1/Q^2_s(Y,b)} \frac{ d r'^2}{r'^2}\,\,+\,\,
\pi\, \int^{r^2}_{1/Q^2_s(Y,b)} \frac{ d |\vec{r} - \vec{r}'|^2}{|\vec{r} - \vec{r}'|^2}
\eeq
The Mellin transform of the full BFKL kernel  has the form
\beq \label{KML}
\chi\Lb \gamma\Rb \,\,=\,\,\int \frac{d \xi}{2 \pi i}\,e^{- \gamma \xi}\,K\Lb r; r'\Rb\,\,=\,\,
2 \psi(1) \,- \,\psi(\gamma) \,-\,\psi(1 - \gamma)
\eeq
where $\xi \,=\,\ln(r^2/r'^2)$ and $\psi(z) = d\ln \Gamma(z)/d z$ with $\Gamma(z)$ equal to Euler gamma function.
The simplified kernel replaces \eq{KML} by the following expression
\bea \label{KSM}
\chi\Lb \gamma\Rb\,\,=\,\,\left\{\begin{array}{l}\,\,\,\frac{1}{\gamma}\,\,\,\,\,\mbox{for}\,\,\,\tau\geq \,1\,;\\ \\
\,\,\,\frac{1}{1 \,-\,\gamma}\,\,\,\,\,\mbox{for}\,\,\,\tau\,\leq\,1\,; \end{array}
\right.
\eea
One can see that the advantage of the simplified kernel of \eq{KSM} is that it provides a matching with the DGLAP evolution equation\cite{DGLAP} in Double Log Approximation (DLA) for $\tau < 1$.

The non-linear BK equation takes two different forms outside and inside the saturation region. For $\tau\,\,<\,\,1$ it can be written as 
\beq \label{BK1}
\frac{\partial^2 n\Lb r,Y; b\Rb}{\partial Y\,\partial \ln\Lb 1/(r^2 \Lambda^2_{QCD})\Rb}\,\,\,=\,\,\frac{\bas}{2}\,\Big( 2 n\Lb r,Y;t=0\Rb\,\,- \,n^2\Lb r,Y; b \Rb\Big)
\eeq
for $n\Lb r,Y; b\Rb\,=\,N\Lb r,Y; b\Rb/r^2$ where $N\Lb r,Y; b\Rb $ is the dipole scattering amplitude.

Inside the saturation region where $\tau\,\,>\,\,1$ the BK equation takes the form
\beq \label{BK2}
\frac{\partial^2 \widetilde{N}\Lb r,Y; b \Rb}{ \partial Y\,\partial \ln r^2}\,\,=\,\, \bas \,\left\{ \Lb 1 \,\,-\,\frac{\partial \widetilde{N}\Lb r,Y; b \Rb}{\partial  \ln r^2}\Rb \, \widetilde{N}\Lb r,Y; b \Rb\right\}
\eeq
where 
 $\widetilde{N}\Lb r,Y; b\Rb\,\,=\,\,\int^{ r^2} d r'^2\,N\Lb r',Y; b \Rb/r'^2$ .
  \subsection{Solution to BK equation}
 Outside the saturation region the non-linear corrections in \eq{BK1} affect the behaviour of the solution to the linear BFKL equation only in the vicinity of the saturation scale  ($\tau \to 1$) where  the solution takes the following form\cite{GSV}
 \beq \label{SOL1}
N\Lb Y; r, b \Rb\,\,\propto \Lb r^2 Q^2_s\Lb Y, b \Rb \Rb^{ 1 - \gamma_{cr}}
\eeq 

where the critical anomalous dimension $\gamma_{cr}$ given by
\beq \label{GACREQ}
- \frac{\partial \omega(\gamma_{cr})}{ \partial \gamma_{cr}}\,\,=\,\,\frac{\omega(\gamma_{cr})}{ 1 - \gamma_{cr}}
\eeq

One can see that \eq{SOL1} shows the geometric scaling behaviour \cite{GS} and it takes the form
\beq \label{SOL10}
N\Lb Y; r, b \Rb\,\, =\,\, N_0 e^{ \h z} ~~~\mbox{where}~~~ z\,\,=\,\,\ln \tau\,\,=\,\,4\,\bas \Lb Y - Y_0\Rb \,+\,\ln\Lb r^2 Q^2_s\Lb Y=Y_0; b\Rb \Rb \,\,=\,\,\xi_s\,\,+\,\xi 
\eeq
since for  the kernel of \eq{KSM} $\gamma_{cr}$=1/2. In  \eq{SOL10} $\xi_s \,=\,4\,\bas \Lb Y - Y_0\Rb$ and $\xi\,=\,
\,\ln\Lb r^2 Q^2_s\Lb Y=Y_0; b \Rb \Rb$. In the entire kinematic region $\tau\,<\,1$ the solution takes the following form for the simplified kernel of \eq{BK1}
\beq \label{SOL11}
N\Lb Y; r, b \Rb\,\,=\,\,N_0\,\exp\Lb \sqrt{-\,\xi_s\,\xi}\,\,+\,\,\xi\Rb \,\xrightarrow{\tau \to 1} \,\,N_0 e^{\h z}\,\exp\Lb - \frac{z^2}{8 \xi_s}\Rb
\eeq
Recall that $N_0 $ is the value of the dipole amplitude at $z=0$ and $\xi < 0$ for $\tau < 1$. One can see that the geometric scaling behaviour holds at $z \,\ll\,8\, \xi_s$.

Solution of \eq{SOL10} provides the boundary condition for the solution inside the saturation region:
\beq \label{INC}
N\Lb Y; \xi = -\xi_s, b \Rb\,\,=\,\,N_0\Lb b \Rb;~~~~~~~~~~~~ \frac{\partial \ln  N\Lb Y; \xi = -\xi_s, b \Rb}{\partial z}\,\,=\,\,\h;
\eeq

Inside the saturation region ($z \,>\,0$) we are looking for the solution in the form\cite{LETU}
\beq \label{NSAT}
\widetilde{N}\,\,=\,\,\int^{\xi}_{\xi_s}
 d \xi'\,\Big( 1\,-\,e^{ - \phi(\xi',Y)}\Big)
\eeq

Substituting \eq{NSAT} into \eq{BK2} we obtain
\beq \label{NSAT1}
 \phi'_Y\,e^{ - \phi}\,\,=\,\, \bas \widetilde{N}\,e^{ - \phi}
\eeq
Canceling $e^{ - \phi}$ and differentiating with respect to $\xi$ we  obtain the equation in the form:
\beq \label{EQXIY}
\frac{\partial^2 \phi}{ \partial Y\,\partial \xi}\,\,=\,\,\,\bas\,\Big( 1\,-\,e^{ - \phi\Lb Y;\xi\Rb}\Big)
\eeq
Using variable $\xi_s$ and $\xi$ we can rewrite \eq{NSAT1} in the form
\beq \label{EQ}
\frac{\partial^2 \phi}{ \partial \xi_s\,\partial \xi}\,\,=\,\,\frac{1}{4}\Big( 1\,-\,e^{ - \phi\Lb Y;\xi\Rb}\Big)~~~\mbox{or in the form}~~~~
\frac{\partial^2 \phi}{ \partial  z^2}\,\,-\,\,\frac{\partial^2 \phi}{ \partial  x^2}\,\,=\,\,\frac{1}{4}\Big( 1\,-\,e^{ - \phi\Lb Y;\xi\Rb}\Big)
\eeq
with  $z$ defined in \eq{SOL10} and $x= \,\xi_s - \xi$.

\eq{EQ} has general traveling wave solution (see Ref.\cite{POL} formula {\bf 3.4.1})

\beq \label{GSOL}
\int^\phi_{\phi_0}\frac{d \phi'}{\sqrt{c \,+\,\frac{1}{2 ( \lambda^2 - \kappa^2)}\Big( \phi'  - 1 + e^{-\phi'}\Big)}}\,\,=\,\,  \kappa \,x +\lambda \,z
\eeq
where $c, \phi_0,\lambda$ and $\kappa$ are arbitrary constants that should be found from the initial and boundary conditions.

From the matching with the perturbative QCD region (see \eq{INC}) we have the following initial conditions for small values of $\phi_0$:
\beq \label{IC}
\phi\Lb t \equiv z = 0, x\Rb\,\,=\,\,\phi_0\Lb b \Rb\,;\,\,\,\,\,\,\phi'_z\Lb t \equiv z = 0, x\Rb\,\,=\,\,\frac{1}{2}\,\phi_0\Lb b \Rb\eeq

These conditions allow us to find that $\kappa=0$ and  $c=0$ for $\phi_0 \,\ll\,1$.
 Therefore, solution of \eq{GSOL} leads to the geometric scaling since it depends only on one variable:   $z$.     \,It takes  the form\cite{LETU,POL} for small values of $\phi_0$
\beq \label{SOLF}
\sqrt{2}\,\int^\phi_{\phi_0}\frac{d \phi'}{\sqrt{ \phi' \,-\,1+ \,e^{-\phi'}}}\,\,=\, \,z
\eeq
\subsection{Generating functional for dilute system}
\begin{figure}
  \leavevmode
  \begin{tabular}{c }
      \includegraphics[width=16cm]{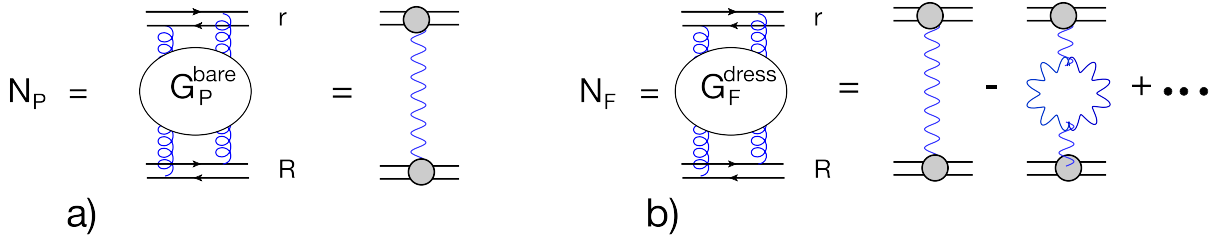}\\
   \end{tabular}
      \caption{ The scattering amplitude of two dipoles with sizes $r$ and $R$.\protect\fig{gpom}-a shows the Green function of the `bare' Pomeron while in \protect\fig{gpom}-b the dressed Pomeron Green function is shown. }
      \label{gpom}
\end{figure}

\eq{SOLF} we can re-write in a different form, namely,
\beq \label{SOLP}
\frac{1}{\sqrt{2}}\,\int^\phi_{\phi_0}d \phi'\Bigg\{ \frac{1}{\sqrt{ \phi' \,-\,1+ \,e^{-\phi'}}}\,\,-\,\,\frac{\sqrt{2}}{\phi'}\Bigg\}\,\,=\,\,\ln N_\pom\Lb \phi_0,z\Rb
\eeq
where 
\beq \label{P}
 N_\pom\Lb \phi_0,z\Rb\,\,=\,\,\frac{\as^2}{16\pi^2}\,G_\pom\Lb \phi_0,z\Rb\,\,=\,\,\phi_0\Lb b \Rb \,e^{\h \,z}
\eeq
where $G_\pom$ is the contribution of one BFKL Pomeron in the saturation region (see \fig{gpom}-a). 
The solution to \eq{SOLP} takes a general form
\beq \label{SOL21}
N\Lb G_\pom\Lb \phi_0,z\Rb\Rb \,\,=\,\,1 \,-\,e^{- \phi(z)}\,\,=\,\,\sum^{\infty}_{n=1} \,\Lb - \,1\Rb^{n+1} C_n\Lb \phi_0\Rb N_\pom^n\Lb \phi_0,z\Rb
\eeq
where $C_n\Lb \phi_0\Rb$ gives the probability to have $n$  BFKL Pomerons at $Y=Y_0$ from the dipole with size $r$ at high energy ($Y$). The sizes of all dipoles delivered by the BFKL Pomerons are equal and about $1/Q_s\Lb Y = Y_0,b \Rb$. $C_n\Lb \phi_0\Rb$ are independent of  the size of the initial dipole due to the geometric scaling behaviour.

  In general the scattering amplitude that we need for the MPSI approximation (see \fig{mpsi}), looks as follows \cite{LMP}:
\beq \label{GEN}
N\Lb Y- Y', r, \{ r_i,b_i\}\Rb\,\,=\,\,\sum^{\infty}_{n=1} \,\Lb - \,1\Rb^{n+1} \widetilde{C}_n\Lb \phi_0, r\Rb \prod^n_{i=1} N_\pom\Lb Y - Y';  r, r_i , b_i\Rb
\eeq
Comparing \eq{GEN} with \eq{SOL21} we see that
\beq \label{CEQC}
\widetilde{C}_n\Lb \phi_0, r\Rb\,\,\,=\,\,\,C_n\Lb \phi_0\Rb
\eeq
 and they do not depend on the size of the initial dipole due to the geometric scaling behaviour of the scattering amplitude. Actually, \eq{CEQC} is the master equation of this paper which will allow us to approach the dipole-dipole scattering amplitude in the MPSI approximation. Proof-as well as a more detailed discussion of this equation, is given in Ref.\cite{LMP}.
 
\begin{figure}
    \begin{tabular}{c c}
  \leavevmode
      \includegraphics[width=8.6cm]{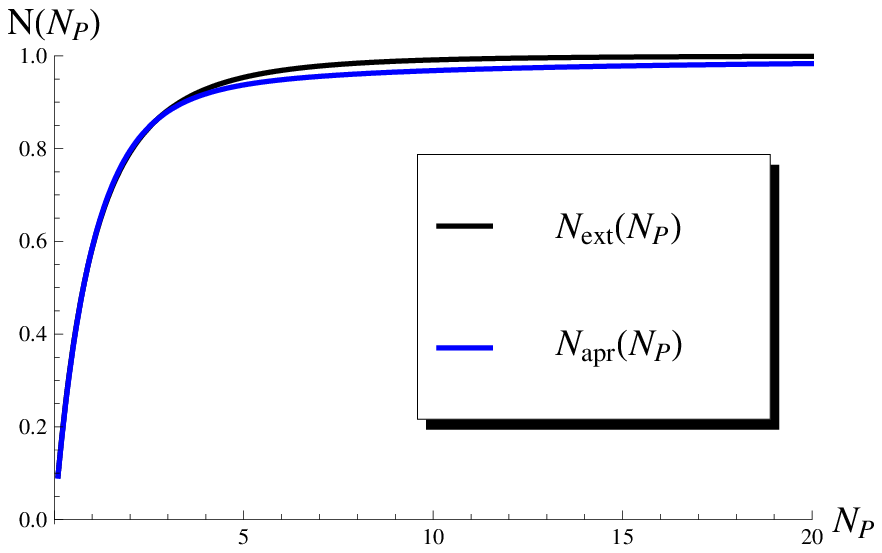}  &     \includegraphics[width=8.6cm]{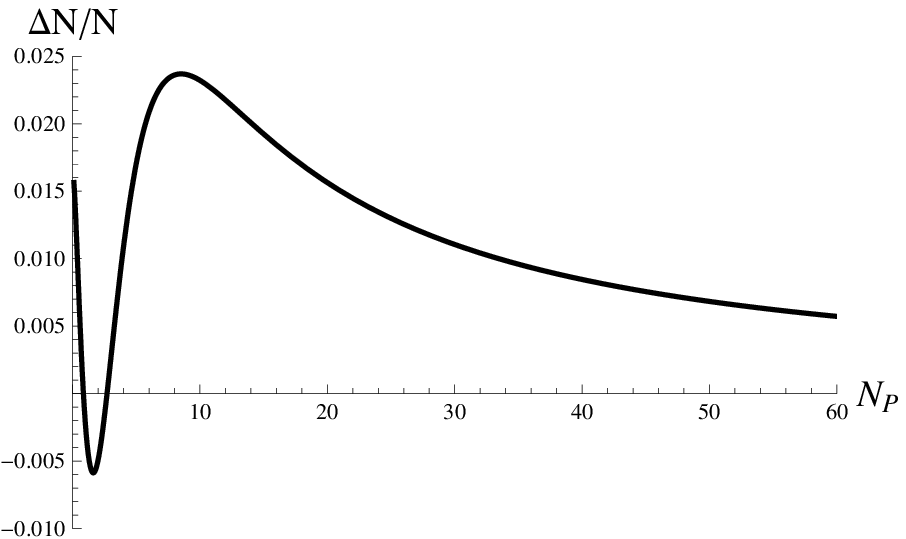} \\
      \fig{sol}-a &\fig{sol}-b\\
            \end{tabular}
\caption{Comparison of the exact solution $N_{ext}\Lb N_\pom\Rb$ and approximate one ($N_{apr}\Lb N_\pom\Rb$ given by  \protect\eq{NAPPR} .  $\Delta N\,\,=\,\,N_{ext}\Lb G_\pom\Rb - N_{apr}\Lb G_\pom\Rb$. \label{sol}}
   \end{figure}

Solving \eq{SOLP} numerically we find $N\Lb G_\pom\Rb$) (see $N_{ext}\Lb G_\pom\Rb$ in \fig{sol}-a). We find that
simple function
\beq \label{NAPPR}
N_{apr}\Lb N_\pom\Rb\,\,=\,\,\kappa \Lb 1 \,-\,\exp\Lb - N_\pom\Rb\Rb\,\,+\,\,\Lb 1 - \kappa\Rb\, \frac{N_\pom}{1\,+\,N_{\pom}}
\eeq
with $\kappa = 0.65$ describes the exact solution within accuracy less that 2.5 \%.( see \fig{sol}-b).
\section{Dressed BFKL Pomeron}

 \eq{GEN} we can find in MPSI approximation \cite{MPSI,LMP,KOLE} the scattering amplitude of two dipoles with sizes $r$ and $R$ and impact parameter $b$ which gives us the Green function of the dressed (final) BFKL Pomeron (see \fig{gpom}-b). It takes the form (see \fig{mpsi})
\bea \label{ADD}
&&\frac{\as^2}{16\,\pi^2}G_F\Lb Y- Y_0, r,  R, b \Rb\,\,\equiv\,N_F\Lb Y - Y_0, r,  R, b \Rb\,\,= \,\,\sum^\infty_{n=1}\,n! \,\Lb - 1\Rb^n\, \Lb \frac{\as}{4 \pi}\Rb^{2 n}\,\\
&& C^2_n \,\prod_{i=1}^n\,\int d^2 r_i d^2 r'_i d^2 b_i, d^2 b'_i \, G^{\mbox{\tiny bare}}_\pom\Lb r, r_i \vec{b} - \vec{b}_i| Y - Y' \Rb \,\gamma\Lb r_i,r'_i,\vec{b}_i - \vec{b'}_i\Rb \,\,G^{\mbox{\tiny bare}}_\pom\Lb r'_i, R, \vec{b'}_i |  Y' \Rb\nn
\eea
\eq{ADD} can be re-written , using \eq{POMTCH} in the form (see \fig{gpom} for notations):
\bea \label{ADDF}
N_{\mbox{dipole-dipole}}\Lb Y- Y_0, r,  R, b \Rb\,\,&= &\,\,N_{F}\Lb Y- Y_0, r,  R, b \Rb\,\,=\,\,\frac{\as^2}{16 \pi^2} G^{\mbox{\tiny dress}}_\pom\,\\
 &=&\,\sum^\infty_{n=1}\,n! \,\Lb - 1\Rb^n\,\Lb \frac{\as}{4 \pi}\Rb^{2 n}\, C^2_n \,\Lb G^{\mbox{\tiny bare}}_\pom\Lb r,  R, b  |  Y - Y_0 \Rb\Rb^n\nn
\eea
From \eq{NAPPR} we derive the approximate simple formula for $N\Lb Y- Y_0, r,  R, b \Rb$ which looks as follows
\bea  \label{ADDAPR}
&&N_{\mbox{dipole-dipole}}^{\mbox{appr}}\Lb Y- Y_0, r,  R, b \Rb\,\,=\,\,N_F^{\mbox{appr}}\Lb Y- Y_0, r,  R, b \Rb \\
&&\kappa^2\Bigg\{ 1 \,-\,\exp\Lb - T\Lb Y- Y_0, r,  R, b \Rb \Rb \,\Bigg\}\,+\,2 \kappa (1 - \kappa) \frac{ T\Lb Y- Y_0, r,  R, b \Rb }{
1\,\,+\,\,T\Lb Y- Y_0, r,  R, b \Rb }\nn\\
  && +\,\,( 1 - \kappa)^2 \,\Bigg\{1 - \exp\Lb \frac{1}{ T\Lb Y- Y_0, r,  R, b \Rb }\Rb\,\frac{1}{T\Lb Y- Y_0, r,  R, b \Rb }\,
  \Gamma\Lb 0,   \frac{1}{
T\Lb Y- Y_0, r,  R, b \Rb }\Rb   \Bigg\}\nn
 \eea 
 
 where $\Gamma\Lb 0,T\Rb $ is incomplete Euler gamma-function (see Ref. \cite{RY} formulae  {\bf 8.35})   
     and 
 \beq \label{T}
 T\Lb Y- Y_0, r,  R, b \Rb \,\,=\,\,  \frac{\as^2}{16 \pi^2}\,G^{\mbox{\tiny bare}}_\pom\Lb r,  R, b  |  Y - Y_0 \Rb\,\,=\,\, 
 N_\pom\Lb Y - Y';  r, R, b \Rb \eeq
 
      \begin{figure}
  \leavevmode
  \begin{center}
      \includegraphics[width=10cm]{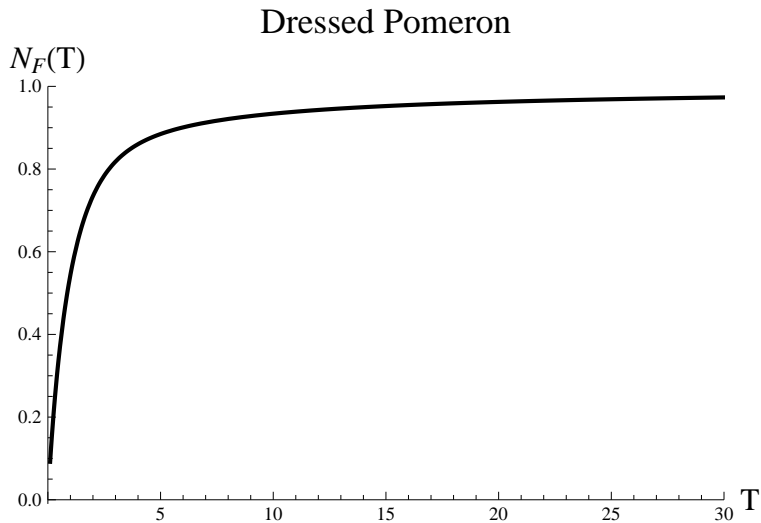}\\
   \end{center}
      \caption{  The dressed Pomeron Green function in the MPSI approximation versus $ T\Lb Y- Y_0, r,  R, b \Rb$.  }
      \label{drpom}
\end{figure} 

Let us discuss the most difficult problem of CGC/saturation approach\cite{KOWI}: the impact parameter dependence of the Pomeron Green Function. For our simplified kernel
 which coincide  with the DGLAP kernel outside of the saturation region,  we can factorize out  the non-perturbative large $b$ behaviour writing for the scattering amplitude in the form:
\beq \label{SA}
 N_\pom\Lb b, Y ,r,R\Rb \,=\,S\Lb b \Rb \int d^2 b'  N^{DGLAP}_\pom\Lb b',  Y ,r\Rb ~~=~~N_0\,r\,R\,S\Lb b \Rb\,e^{\h z} 
 \eeq
 where $N_0$ is a constant
  (see Ref. \cite{GLLMT}).  In our estimates we choose $r =R = 1/m$.
  
\begin{figure}
  \leavevmode
  \begin{center}
      \includegraphics[width=6cm]{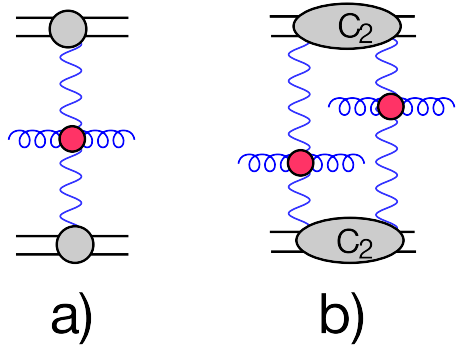}\\
   \end{center}
      \caption{  Double parton interaction due to  Pomeron loops.   }
      \label{cor}
\end{figure}

  Indeed, considering the scattering amplitude at fixed  transferred  momentum $q $ ( which is Fourier conjugated to $b$), one can see that for $q \,< \,\mu_{soft}$ the evolutions in $ln(1/r)$ do not depend on $q$. However, for $q\, > \,\mu_{soft}$ the logs take the form $\ln\Lb1/\Lb r q \Rb\Rb$ and the $q$ dependence cannot be absorbed in $S\Lb b \Rb$ in \eq{SA}\cite{GLLMT}.  Using \eq{SA} we can absorbed the non-perturbative corrections at large $b$ in the definition of the saturation scale $Q_s\Lb Y; b\Rb$\cite{MU90,LETU,BRAUN,ARBR,LGLM}. $S\Lb b \Rb$ is a non-perturbative form factor which we parameterize in the form
  \beq \label{S}
  S\Lb b \Rb\,\,=\,\,\frac{m^2}{2 \pi} \,e^{- m\,b}\,;~~~~~~~~~~\int d^2 b\,S\Lb b \Rb\,\,=\,\,1\,;
  \eeq
  introducing mass $m$. Using the experimental data on cross section off the double parton interaction (see \eq{XSEFF}) we can write that (see \fig{cor})
  \beq \label{m}
2\, C^2_2\, \frac{m^2}{8 \pi}\,=\,\frac{1}{ \sigma_{eff}}
 \eeq
 \eq{m} is written assuming that only Pomeron loops contribute to the double parton interaction (see \fig{cor}). This assumption looks natural in CGC/saturation approach for the proton-proton scattering as has been discussed in the introduction but, being phenomenological, it should be re-check in the future description of proton-proton date based on the result of this paper. Plugging the experimental value $\sigma_{eff}\,=\,12\,\div\,15 \,mb$ we obtain
 $m\,=\,0.86\,\div\,1\,GeV$. This value of $m$ is in a good agreement with other indications   of the second dimensional scale in the proton \cite{KPPS}. The variable $z$ in \eq{SA} is defined as $ z\,\,=\,\,4 \bas\,\Lb Y - Y_0\Rb \,+\,\ln\Lb r^2/R^2\Rb$. Collecting \eq{SA} and \eq{S} the Pomeron Green function takes the following form
 \beq \label{PGEN}
  N_\pom\Lb b, Y ,r,R\Rb \,=\, r\,R\,\,N_0\,S\Lb b \Rb\,e^{2 \bas \Lb Y - Y_0\Rb\,+\,2 \ln\Lb r^2/R^2\Rb}
  \eeq
  
  In all our numerical estimates we take $r=R=1/m$ and $N_0=1/2$. It should be mentioned that any other choice will lead only to redefinition of the value for $N_0$ in \eq{PGEN}. With our choice we have
  \beq \label{NFIN}
  N_\pom\Lb b, Y ,r,R\Rb \,=\, \,0.08\, e^{ - m\,b}\,e^{2 \bas \Lb Y - Y_0\Rb\,+\,2 \ln\Lb r^2/R^2\Rb}
  \eeq  
  Certainly $N_\pom \Lb b =0, z=0,\Rb = 0.08\,\,\ll\,\,1$ satisfies the condition, that we have used 
  obtaining the solution to the non-linear equation.

\section{Diffractive production in dipole-dense target interaction: solution to the equation}

\begin{figure}
  \leavevmode
  \begin{center}
      \includegraphics[width=14cm]{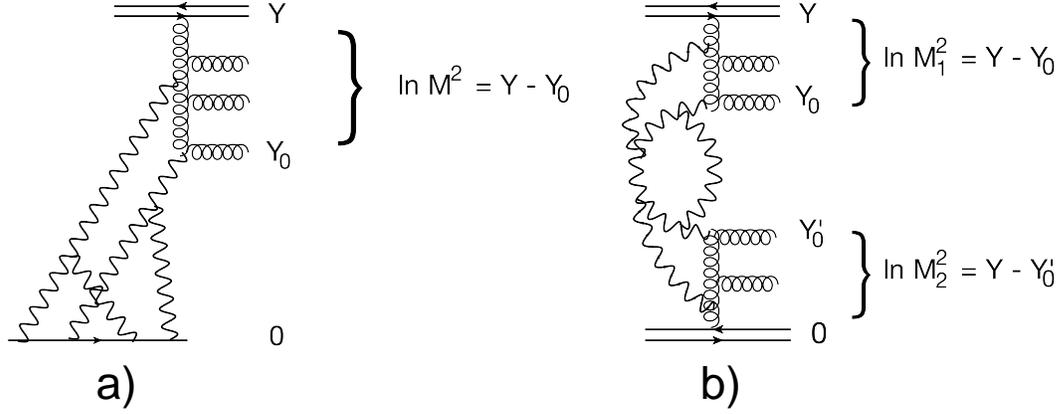}\\
   \end{center}
      \caption{  Example of the diagrams for the diffractive production:single diffraction ( \protect\fig{dif}-a) and double diffraction( \protect\fig{dif}-b). Helix lines denote gluons, wavy lines describe the BFKL Pomeron.  }
      \label{dif}
\end{figure}

The equation for the diffractive production in dipole-target scattering (see \fig{dif}) has been known for more than decade\cite{KOLEV} (see also Ref.\cite{REV}). It has the same form as BK equation but for the following amplitude
\beq \label{DEF}
G\Lb Y; Y_0, r,b \Rb\,\,=\,\,2 N\Lb Y, r, b \Rb \,-\,N^D\Lb Y,Y_0,r, b \Rb
\eeq
The  cross section of diffractive production with the rapidity gap ($Y - \ln M^2$  in \fig{dif}) larger than $Y_0$
can be written through $N^D\Lb Y,Y_0,r, b \Rb $ :
\beq \label{D1}
\sigma_{diff}\,\,=\,\,\int d^2 b\,  N^D\Lb Y,Y_0,r, b \Rb
\eeq
From \eq{D1}and unitarity constraint  follows that $G\Lb Y; Y_0, r,b \Rb$ is the inelastic cross section cross section of all processes except the diffractive production with the rapidity gap $\geq\,Y_0$. At $Y= Y_0$
\beq \label{DEFI}
G\Lb Y_0; Y_0, r,b \Rb\,\,=\,\,2 N\Lb Y_0, r, b \Rb \,-\,N^2\Lb Y,Y_0,r, b \Rb
\eeq
where $N^2$ is the elastic cross section. \eq{DEFI} determines the initial condition for the equation for
$G\Lb Y; Y_0, r,b \Rb$. Since the equation for $G$ is the same as for $N$, therefore, the solution in the saturation region is equal to 
$G\Lb z,z_0\Rb\,\,=\,\, 1 - \exp\Lb - \phi\Lb z \Rb\Rb$ where $\phi$ is given by \eq{GSOL}. The difference is the initial condition is given by  \eq{DEFI} at $Y = Y_0$. Here, we consider the case when both  $Y$ and  $Y_0$ are so large that we have the geometric scaling behaviour for the amplitude both at $Y_0$ and $Y$, We introduce two variables
\beq \label{2Z}
z\,\,=\,\,4 \bas Y + \ln \Lb r^2/R^2\Rb;~~~~~~~~~z_0\,\,=\,\,4 \bas Y_0 + \ln \Lb r'^2/R^2\Rb
\eeq
where $r$ and $r'$ are the sizes of the dipoles with rapidities $Y$ and $Y_0$, respectively.

Rewriting the  initial conditions in the variables of \eq{2Z} we have
\beq \label{ICD}
G\Lb z_0,z_0; b \Rb\,\,=\,\,1\,-\,\exp\Lb - \phi\Lb \Delta z \Rb \Rb\,\,=\,\,1\,-\,\exp\Lb -2 \phi\Lb z_0; b \Rb\Rb;~~~\mbox{or}~~~~ \phi\Lb \Delta z(z_0; b )\Rb\,=\,2 \,\phi\Lb z_0; b \Rb
\eeq
Final solution takes the form
\beq \label{SOLD}
G\Lb z,z_0; b\Rb\,\,=\,\,1\, - \, \exp\Big( - \phi\Lb z - z_0 + \Delta\Lb z_0; b\Rb\; b\Rb\Big)
\eeq
and 
\beq \label{SOLD1}
N^D\Lb z,z_0; b \Rb\,\,=\,\,1\, -\,2\,\exp\Big( - \phi\Lb z; b \Rb\Big)\,+\,\exp\Big( - \phi\Lb z - z_0 + \Delta\Lb z_0; b \Rb, b\Rb\Big)
\eeq

\begin{figure}
  \leavevmode
  \begin{center}
      \includegraphics[width=10cm]{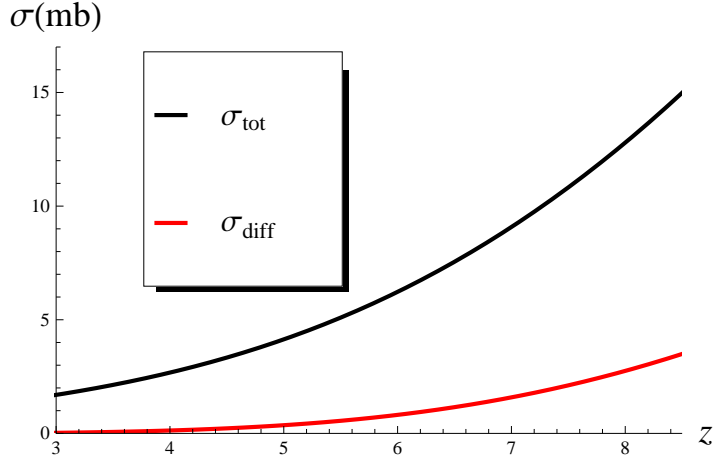}\\
   \end{center}
      \caption{   The total cross section and the cross section of the diffraction production with $z_0 = 3$. In  \protect\eq{PGEN} $\bas = 0.2$,$N_0$ = 0.5 }
      \label{stdiff}
\end{figure}

  In \fig{stdiff} we plot the total    and diffraction cross sections. We fix $N_{pom}$ in \eq{PGEN} considering $N_0 = 0.5$.  Using this form of $N_\pom$ and solution of \eq{SOLP} we can calculate
  \bea \label{XS}
  \sigma_{tot}\,\,&=&\,\,2 \int d^2 b \Big( 1 \,\,-\,\,\exp\Lb - \phi\Lb N_\pom\Lb z, b \Rb\Rb\Rb\Big)\,;\nn\\
  \sigma_{diff}\,\,&=&\,\,\int d^2 b \Big( 1 \,\,-\,\,2\,\exp\Lb - \phi\Lb N_\pom\Lb z, b \Rb\Rb\Rb\,  +\,\exp\Lb - \phi\Lb
   N_{\pom}\Lb z - z_0 + \Delta\Lb z_0, b\Rb; b \Rb\Rb\Rb\Big)  
  \eea
  
\begin{figure}
    \begin{tabular}{c c}
  \leavevmode
      \includegraphics[width=8.6cm]{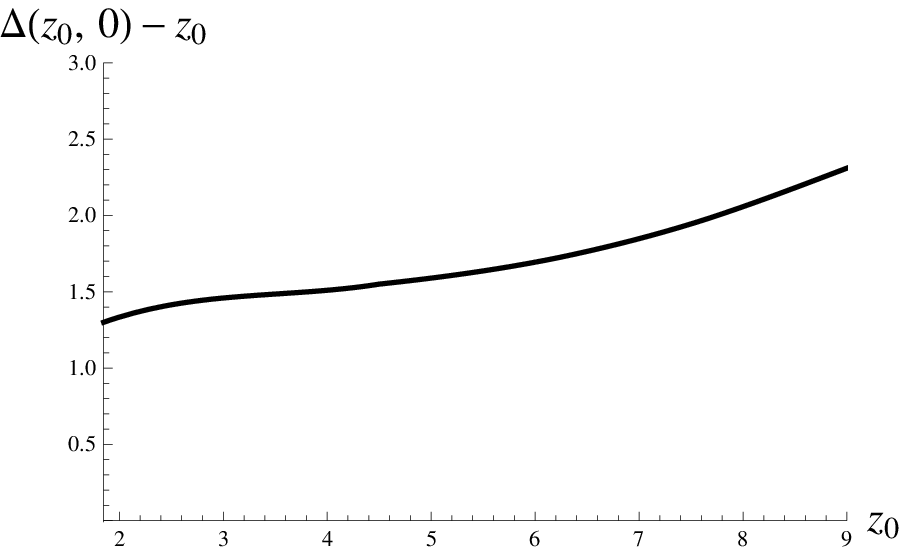}  &     \includegraphics[width=8.6cm]{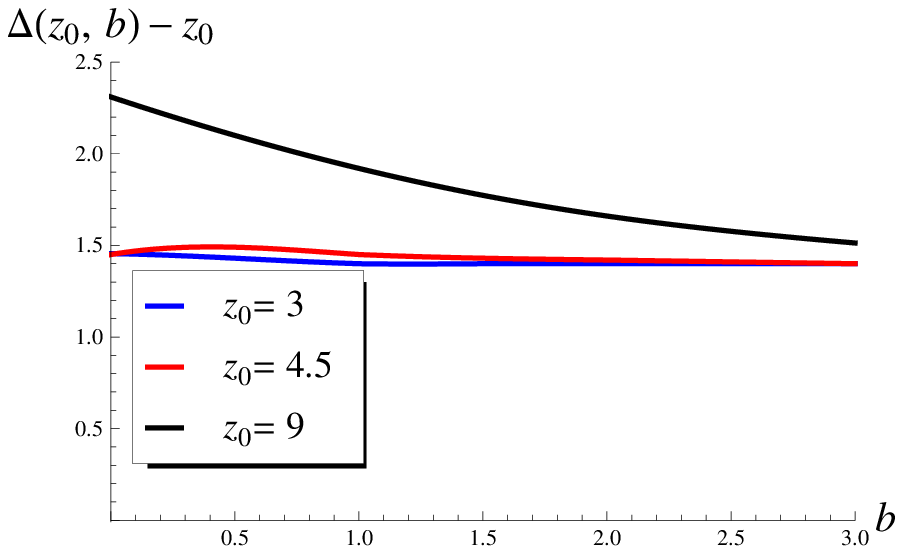} \\
      \fig{delta}-a &\fig{delta}-b\\
            \end{tabular}
\caption{Behaviour of $\Delta\Lb z_0, b\Rb$ versus $z_0$ (  \protect\fig{delta}-a) and versus $b$ ( \protect\fig{delta}-b). \label{delta}}
   \end{figure}

  In \fig{delta}  we plot  $\Delta\Lb z_0,b\Rb - z_0$ versus $z_0$ and $b$. One can see that at small $z_0$ \eq{ICD} can be re-written in the form
  \beq \label{SOLD2}
   \phi\Lb \Delta z(z_0)\Rb\,\,=\,2 \,\phi\Lb N_\pom\Lb z_0,b\Rb \Rb\,\,\xrightarrow{ N_\pom \ll 1}\,\, \,\phi\Lb 2 N_\pom\Lb z_0,b\Rb \Rb  
\eeq
leading to $\Delta\Lb z_0\Rb - z_0 \,=\,2 \ln 2$ for $\bas = 0.25$. From \eq{XS} we can calculate $M^2 d \sigma_{diff}/d \ln M^2$ (see \fig{dif}) which takes the form
\beq \label{SOLD3}
\frac{d \sigma_{diff}}{d \ln M^2}\,\,=\,\,-\,\frac{d N^D}{d Y_0}\,\,=\,\,-\,4\,\bas\,\frac{d N^D}{d z_0}\,\,=\,\,2\bas \frac{ d \phi\Lb N_\pom\Rb}{ d \ln N_\pom}\, \frac{d \Lb \Delta\Lb z_0, b \Rb - z_0\Rb}{d z_0}\,\,e^{  - \phi\Lb N_\pom\Lb z -z_0 +  \Delta\Lb z_0, b\Rb , b \Rb\Rb}
\eeq

\begin{figure}
        \begin{tabular}{c c}
          \leavevmode
      \includegraphics[width=8cm]{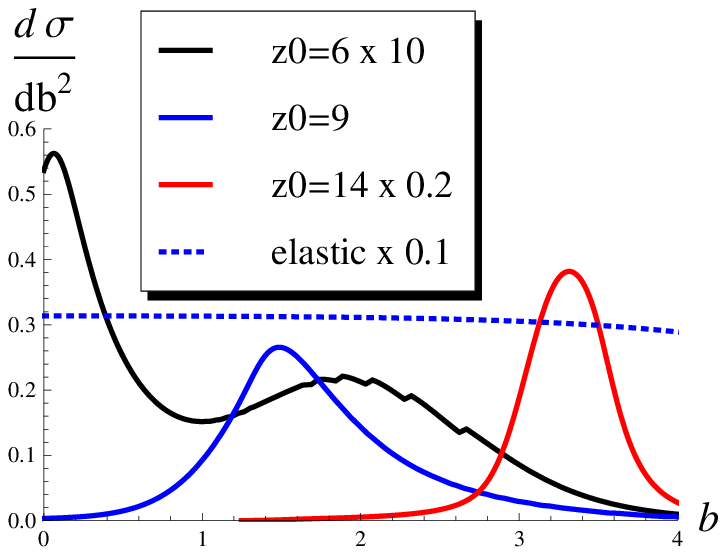} &  \includegraphics[width=8cm]{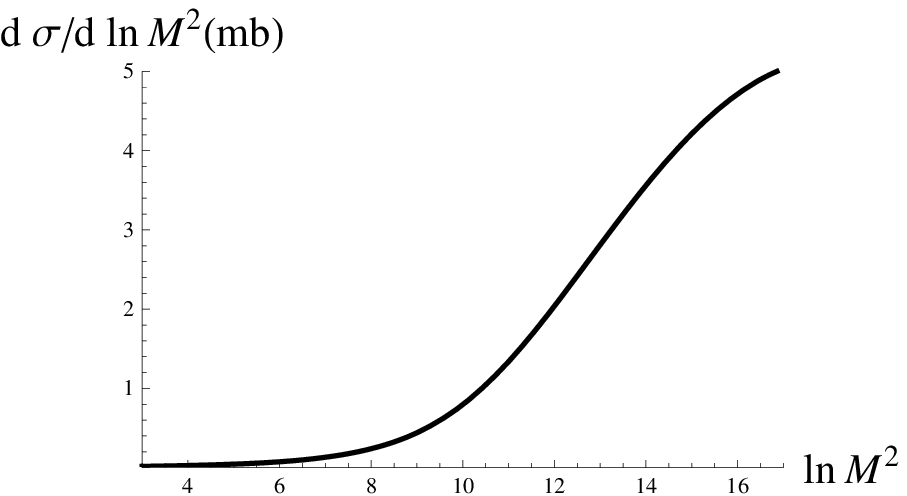}\\
      \fig{sdfixm}-a &     \fig{sdfixm}-b\\
                  \end{tabular}
\caption{ \protect\fig{sdfixm}-a shows the impact parameter dependence of the single diffraction at different values of $z_0$ for fixed $z=17.75$. The curves  are multiplied by factors:  at $z_0 = 6$ by 0.1 and at $ z_0=14$ by 0.2 .  For elastic cross section the factor is 0.1.
Both $\frac{d \sigma_{diff}}{d \ln M^2\,d b }  $ and $b$ are shown in $1/GeV$. In  \protect\fig{sdfixm}-b it is plotted the dependence of the diffraction cross section at fixed mass ($d \sigma_{diff}/d \ln M$) at $z=17.75$ which corresponds to the LHC energy for $\bas = 0.25$. }
\label{sdfixm}
   \end{figure}

In \fig{sdfixm} we plot  $\frac{d \sigma_{diff}}{d \ln M^2\,d b }  $ at fixed $z =17.75$ which for $\bas=0.25$ corresponds to the LHC energy. One can see that the largest contribution stems from the large  mass kinematic region and the $b$-dependence  shows a peripheral  - type of behaviour versus $b$ with maximum at $b \approx 6$. We plot in this figure the elastic cross section at fixed $b$ ( $A^2_{el}(b)$) to illustrate the peripheral character of the diffraction production.

\section{ Single diffractive production for  dipole-dipole interaction in the MPSI approximation}

\begin{figure}
  \leavevmode
  \begin{center}
      \includegraphics[width=14cm]{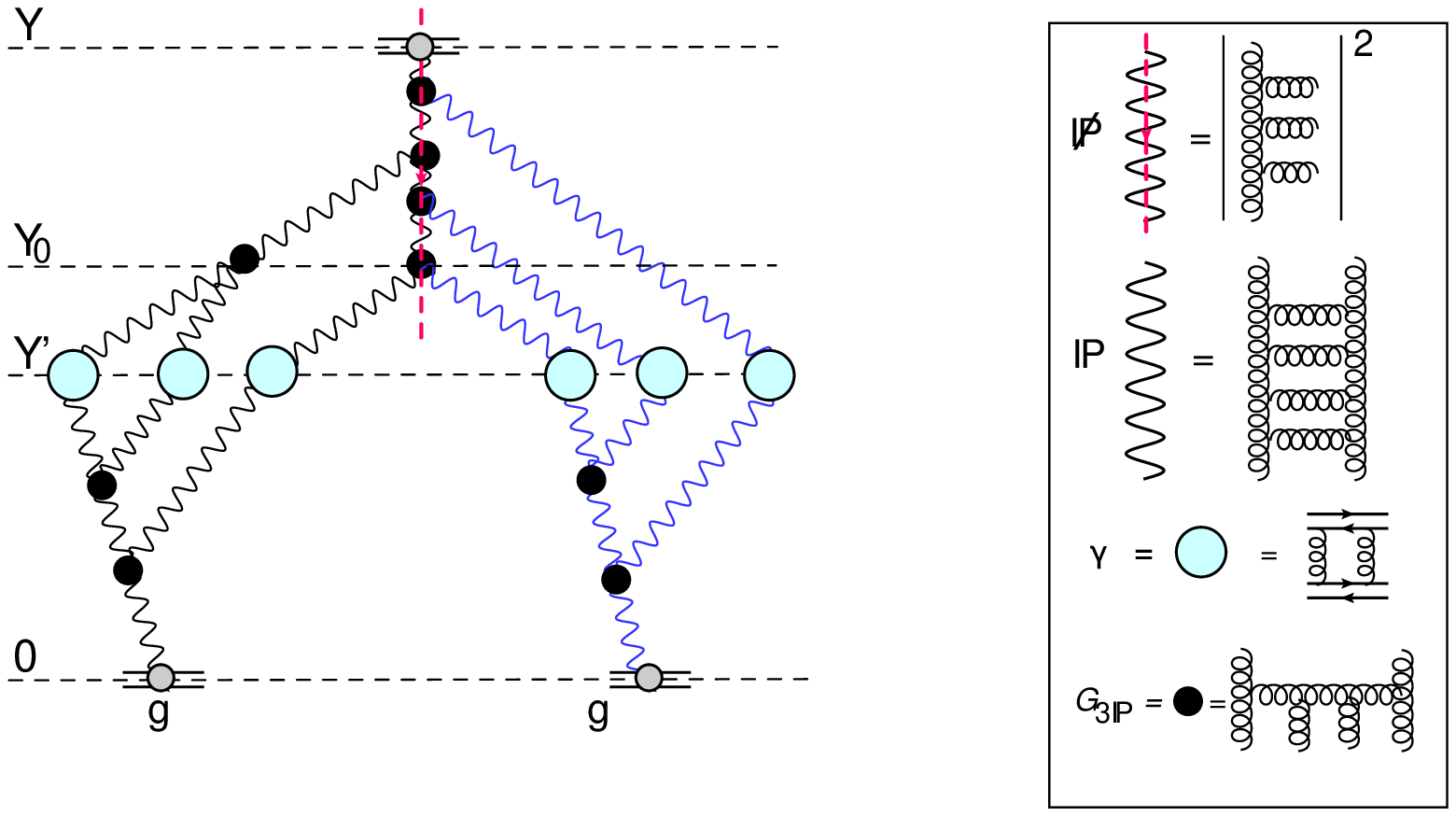}
   \end{center}
      \caption{ MPSI approximation for diffraction production:  all notations are shown in the insertion. Wavy lines denote the BFKL Pomerons, 
      the blob stands for the scattering amplitude of two dipoles with sizes:$ r_1$ and $r_2$ in the Born approximation (due to exchange of two gluons). $G_{3 \pom}$ is the triple BFKL Pomerons vertex. $\pom\!\!\!\!\!\Big{\slash}$ denotes the cut Pomeron shown in the insertion.}
      \label{mpsisd}
\end{figure}

The pattern of calculation of diffractive production in the dipole-dipole scattering is shown in \fig{mpsisd}.This picture illustrates the main difference between calculation of the scattering amplitude and the cross section of the diffractive production: for the latter we need to introduce the difference between BFKL Pomerons in the scattering amplitude (in black in \fig{mpsisd} and in the complex conjugated amplitude shown in blue in \fig{mpsisd} (see Ref. \cite{REV, KLP} and references therein).  The general equation for the cross section of the single diffractive production takes the form

\bea \label{SDDD}
&&N^D_{\mbox{dipole-dipole}}\,\,=\,\,N^D\Lb z -  z',z_0, \vec{b} - \vec{b}^{'}\Rb \bigotimes N \Lb z', b'\Rb N^{*} \Lb z', b'\Rb\,\,=\,\, \,\sum^\infty_{n=1}  \sum_{k=1}^{n - 1}\,n!\Big(\frac{(n - k)! k!}{n !}\Big)^2 \,\Lb - 1\Rb^n\,\nn\\
&&\times \Lb \frac{\as}{4 \pi}\Rb^{2 n}\, C^D_{n-k,k} C_{n -k} \,C_{k}\,\Big( G^{\mbox{\tiny bare}}_\pom\Lb z \,+\,\Delta\Lb z_0, b\Rb - z_0; b\Rb\Big)^{n -k} \Big( \widetilde{G}^{\mbox{\tiny bare}}_\pom\Lb z \,+\,\Delta\Lb z_0, b\Rb - z_0; b\Rb\Big)^{k}
\eea
where $\widetilde{G}$ denotes the Pomeron Green's function in the complex conjugated amplitude ($N^{*}$) and  $C^D_{n - k, k}$ and $C_n$ are the coefficients in the series:
\bea\label{SDDD1}
N^{D}\Lb z -  z', z_0; b\Rb \,\,&=&\,\,\sum_{n=1}^{\infty}\sum_{k=1}^{n-1} C^D_{n-k. k} \Lb G^{\mbox{\tiny bare}}_\pom\Lb  z- z'  - z_0 + \Delta\Lb z_0, b\Rb \Rb; b \Rb^{n - k}\Lb \widetilde{G}^{\mbox{\tiny bare}}_\pom\Lb  z- z'  - z_0 + \Delta\Lb z_0, b\Rb \Rb; b \Rb^{k}\nn\\
N\Lb z'\Rb &=& \sum_{n=1}^{\infty} C_n \Lb G^{\mbox{\tiny bare}}_\pom\Lb  z', b \Rb\Rb^n
\eea

As has been shown in Ref.\cite{KLP} we need to re-write \eq{SOLD} and \eq{SOLD1} in the following form  to obtain coefficients $
C^D_{n - k, k}$
\bea
&&G\Lb G^{\mbox{\tiny bare}}_\pom\Lb z - z';b\Rb, \widetilde{G}^{\mbox{\tiny bare}}_\pom\Lb z - z'\; b\Rb\Rb\,\,= \label{SDDD2}\\
&&1\, - \, \exp\Bigg( - \phi\Big( e^{\Delta\Lb z_0; b\Rb - \ln 2} \Big\{G^{\mbox{\tiny bare}}_\pom\Lb  z- z'  - z_0 ; b \Rb  +  \widetilde{G}^{\mbox{\tiny bare}}_\pom\Lb  z- z'  - z_0 ; b \Rb\Big\}\Big)\Bigg) \nn\\
&&N^D\Lb z- z',z_0; b \Rb\,\,=\label{SDDD3}\\
&&N \Lb G^{\mbox{\tiny bare}}_\pom\Lb z - z';b\Rb\Rb\,\,+\,\,
N \Lb \widetilde{G}^{\mbox{\tiny bare}}_\pom\Lb z - z';b\Rb\Rb\,-\,G\Lb G^{\mbox{\tiny bare}}_\pom\Lb z - z';b\Rb, \widetilde{G}^{\mbox{\tiny bare}}_\pom\Lb z - z'\; b\Rb\Rb\nn
\eea

Using \eq{SDDD2} and \eq{SDDD3} we can simplify \eq{SDDD1} reducing it to the form:
\bea\label{SDDD4}
&&N^{D}\Lb z -  z', z_0; b\Rb \,\,=\\
&&\sum_{n=1}^{\infty}\sum_{k=1}^{n-1}\frac{n!}{(n - k)! k!} C^D_{n} 
\Lb G^{\mbox{\tiny bare}}_\pom\Lb  z- z'  - z_0 + \Delta\Lb z_0, b\Rb \Rb; b \Rb^{n - k}\Lb \widetilde{G}^{\mbox{\tiny bare}}_\pom\Lb  z- z'  - z_0 + \Delta\Lb z_0, b\Rb \Rb; b \Rb^{k}\nn
\eea
leading to $C^D_{n - k,k} \,\,=\,\,\frac{n!}{(n - k)! k!} C^D_{n} $. Note, that two first terms in \eq{SDDD3} do not contribute to \eq{SDDD}.

Performing summation over $n$ and $k$ in \eq{SDDD1} we use for $N$ the approximate expression of \eq{NAPPR} which  we re-write in the following form
\beq \label{TRICK}
N_{appr}\Lb T^{\mbox{\tiny bare}}_\pom\Rb \,\,=\,\,\int^\infty_0 d t e^{-t}\Big( 1\,\,-\,\,\kappa e^{ - T^{\mbox{\tiny bare}}_\pom}\,\,-\,(1 - \kappa)\,e^{ - t \,T^{\mbox{\tiny bare}}_\pom}\Big)
\eeq
The result of lengthy but simple calculation takes the form

\bea \label{SDDDF}
N^D_{\mbox{\tiny dipole-dipole}}\Lb z, z_0,b\Rb \,\,&=& \kappa\Big( \kappa\Lb 1 - \exp\Lb - {\cal T}\Rb\Rb\,+\,(1 - \kappa)\frac{{\cal T}}{1 \,+\,{\cal T}}\Big)^2 \,+\,\kappa^2 (1 - \kappa)\frac{2 {\cal T}^2}{\Lb 1\,+\,{\cal T}\Rb\, \Lb 1 \,+\,2 {\cal T}\Rb}\nn\\
&+&\,\kappa (1 - \kappa)^2\Big( \frac{{\cal T}}{1 + {\cal T}}\,+\,\exp\Lb 1 + \frac{1}{\cal T}\Rb \Gamma\Lb 0, 1 + \frac{1}{\cal T}\Rb\,-\,\exp\Lb \frac{1}{\cal T}\Rb \Gamma\Lb 0,  \frac{1}{\cal T}\Rb\Big)\nn\\
 &+& (1 - \kappa)^3  \frac{1}{{\cal T}^2}\Big( {\cal T}\Lb 1 + {\cal T}\Rb \,-\,\exp\Lb \frac{1}{\cal T}\Rb
 \Lb 1 + 2 {\cal T}\Rb \Gamma\Lb 0,  \frac{1}{\cal T}\Rb\Big)
 \eea
 where 
 \beq \label{CALT}
 {\cal T}\,\,\equiv\,\,\frac{\as^2}{16 \pi^2}\,e^{2 \bas ( \Delta\Lb z_0,b\Rb -2 \ln 2)}G^{\mbox{\tiny bare}}_\pom\Lb z; b \Rb
  \eeq
 
     \begin{figure}
    \begin{tabular}{c c}
  \leavevmode
      \includegraphics[width=8.6cm]{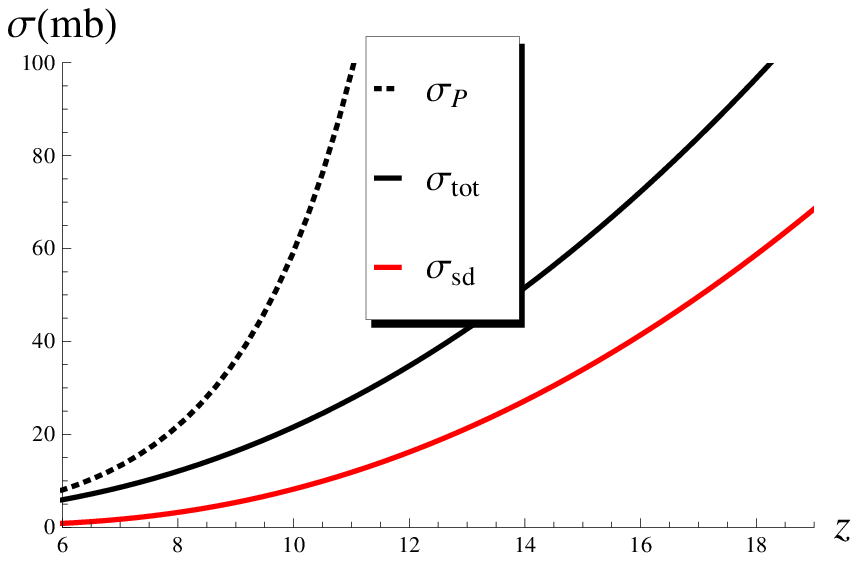}  &     \includegraphics[width=8.6cm]{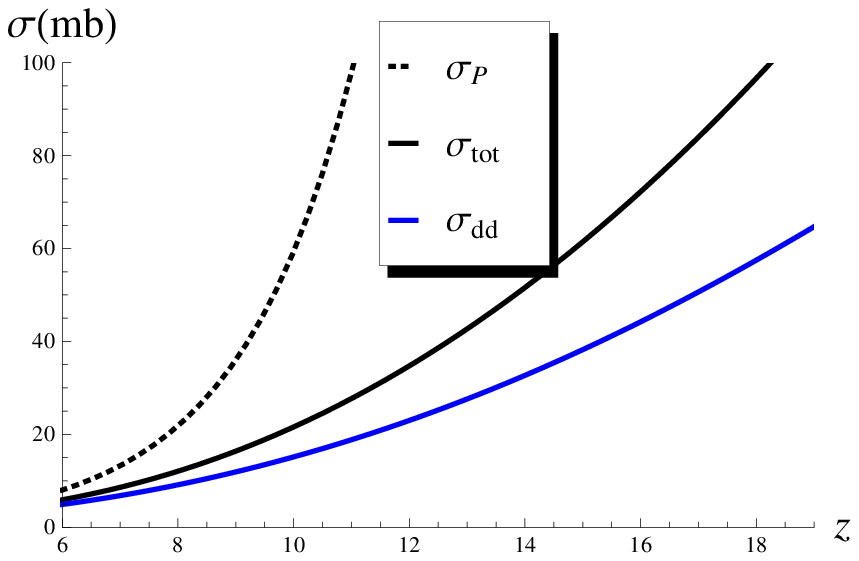} \\
      \fig{diff}-a &\fig{diff}-b\\
            \end{tabular}
\caption{Total and diffractive production cross sections for the dipole-dipole scattering versus $z$. For single ($\sigma_{sd}$) and for double ($\sigma_{dd}$) diffraction the minimal rapidity gap is chosen $ Y_0 = 3$, $\bas = 0.25$. The dotted line shows the contribution of the `bare' Pomeron exchange to the total cross section.  The value of $N_0$ and the mass in $S\Lb b\Rb$ in  \protect\eq{PGEN} are chosen: $N_0 = 0.5$ and $m = 1\,GeV$.
 \label{diff}}
   \end{figure}
In \fig{diff} we present the result of our estimates for total, single and double diffraction cross sections.
One can see that the single diffraction is not small reaching about  a quarter of the total cross sections at large values of $z$.


\section{ Double  diffractive production for  dipole-dipole interaction in the MPSI approximation}
\begin{figure}
  \leavevmode
  \begin{center}
      \includegraphics[width=14cm]{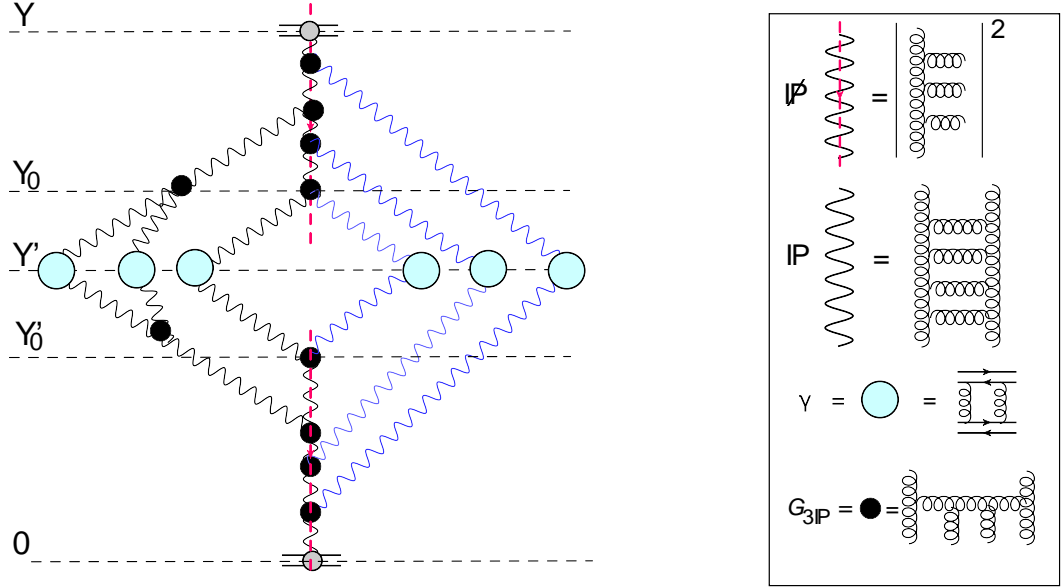}
   \end{center}
      \caption{ MPSI approximation for double diffraction production:  all notations are shown in the insertion. Wavy lines denote the BFKL Pomerons, 
      the blob stands for the scattering amplitude of two dipoles with sizes:$ r_1$ and $r_2$ in the Born approximation (due to exchange of two gluons). $G_{3 \pom}$ is the triple BFKL Pomerons vertex. $\pom\!\!\!\!\!\Big{\slash}$ denotes the cut Pomeron shown in the insertion.}
      \label{mpsidd}
\end{figure}

We can calculate in the framework of the MPSI approximation the cross section of the double diffractive production (see \fig{dif}-b and \fig{mpsidd})) using the following equations
\bea \label{DDD}
&&N^{DD}_{\mbox{dipole-dipole}}\,\,=\\
&&\,\,N^D\Lb z -  z',z_0, \vec{b} - \vec{b}^{'}\Rb \bigotimes N^D \Lb z', z'_0b'\Rb =\,\, \,\sum^\infty_{n=1}  \sum_{k=1}^{n - 1}\,n!\Big(\frac{(n - k)! k!}{n !}\Big)^2 \,\Lb - 1\Rb^n\,
 \Lb \frac{\as}{4 \pi}\Rb^{2 n}\Lb C^D_{n-k,k}\Rb^2\nn\\
 &&\times\, \Big( G^{\mbox{\tiny bare}}_\pom\Lb z \,+\,\Delta\Lb z_0, b\Rb +\Delta\Lb z'_0, b\Rb  - z_0  -z'_0; b\Rb\Big)^{n -k} \Big( \widetilde{G}^{\mbox{\tiny bare}}_\pom\Lb z \,+\,\Delta\Lb z_0, b\Rb \,+\Delta\Lb z'_0, b\Rb \,- z_0\,-\,z'_0; b\Rb\Big)^{k}\nn
\eea

Substituting in \eq{DDD} \eq{SDDD2} and \eq{SDDD3} we obtain a very economic expression for $N^{DD}_{\mbox{dipole-dipole}}$, viz.
\beq \label{DDD1}
N^{DD}_{\mbox{dipole-dipole}}\Lb z, z_0,z'_0; b\Rb\,\,\,=\,\,N_{\mbox{\tiny dipole-dipole}}\Lb z + \Delta\Lb z_0;b\Rb + \Delta\Lb z'_0; b \Rb - 2 \ln 2 - z_0 - z'_0 \Rb
\eeq
\fig{diff} demonstrates that the cross section for the double diffractive production is close to the cross sections of the single diffractive production and its value can be rather large.

\section{Density of the produced gluons}

\subsection{Single inclusive production}


Armed  with the knowledge of the dipole-dipole scattering amplitude we are ready to answer what kind of parton system is produced at high energy. First, we will find the density of the produced gluons per unit of rapidity ($d N_G/dY_1$)  which is equal to \cite{KOTU} (see also Refs.\cite{REV,LERE} for details).
\bea \label{GD}
&&\frac{d N_G}{d Y_1} \,=\,\int d^2 p_T\frac{1}{\sigma_{in}}\frac{d \sigma}{d y \,d^2 p_{T}}\,\,=\\
&& \frac{2C_F}{\bas (2 \pi)^4 }\,\int d^2 p_T\frac{1}{p^2_T}\int d^2 b \,d^2 b' \,d^2 r_\perp\,e^{i \vec{p}_T\cdot \vec{r}_\perp}\,\,\nabla^2_\perp\,N^{G}_{\mbox{\tiny dipole-dipole}}\Lb Y_1 ; r_\perp; b \Rb\,\,\nabla^2_\perp\,N^{G}_{\mbox{\tiny dipole-dipole}}\Lb Y_1; r_\perp; b' \Rb\nn
\eea
where \cite{KOTU}
\beq \label{GD1}
N^{G}_{\mbox{\tiny dipole-dipole}}\Lb Y_1 ; r_\perp; b \Rb\,\,=\,\,2\,N_{\mbox{\tiny dipole-dipole}}\Lb Y_1 ; r_\perp; b \Rb \,-\,N^2_{\mbox{\tiny dipole-dipole}}\Lb Y_1 ; r_\perp; b \Rb
\eeq
Integration over $r_\perp$ in \eq{GD} spans  over saturation region ( $z > 0$) as well as over perturbative QCD domain ($z < 0$). Since we calculated $N_{\mbox{\tiny dipole-dipole}}$ in in the saturation region we can claim that
\beq \label{GD2}
\frac{d N_G}{d Y_1}\,\geq \,\frac{1}{\sigma_{in}}\frac{2C_F}{\bas (2 \pi)^4 }\,\int d^2 p_T\frac{1}{p^2_T}\int d^2 b \,d^2 b' \,\int_{1/Q_s(Y_1)}\!\!\!\!\!\!\!\!\!\!\!\!\!\!\!\!\!\!\pi d r^2_\perp\,e^{i \vec{p}_T\cdot \vec{r}_\perp}\,\,\nabla^2_\perp\,N^{G}_{\mbox{\tiny dipole-dipole}}\Lb Y_1 ; r_\perp; b \Rb\,\,\nabla^2_\perp\,N^{G}_{\mbox{\tiny dipole-dipole}}\Lb Y_1; r_\perp; b' \Rb
\eeq
which in the notation of \eq{SOL10} can be re-written in the form
\beq \label{GD3}
\frac{d N_G}{d Y_1}\,\geq \,\frac{1}{\sigma_{in}}\frac{2C_F}{\bas  \pi^2 }\,\int d l \int d^2 b \,d^2 b' \,\int_{-\xi_s}^{\infty} d \xi e^{-\xi}J_0\Lb e^{\h l +\h \xi}\Rb\,\frac{d^2}{d  \xi^2}N^{G}_{\mbox{\tiny dipole-dipole}}\Lb \xi_s + \xi, b \Rb\,N^{G}_{\mbox{\tiny dipole-dipole}}\Lb \xi_s + \xi, b' \Rb\
\eeq
where $l = \ln p^2_T$.

  \begin{figure}
  \leavevmode
  \begin{center}
      \includegraphics[width=3cm]{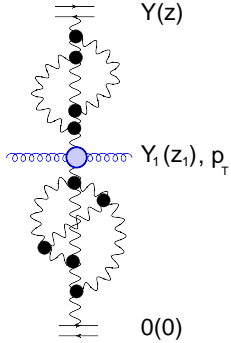}
   \end{center}
      \caption{ Mueller diagram for the inclusive production of gluons with rapidity $Y_1$ and $z_1$ and with transverse momentum $p_T$. All notations are the same as in \protect\fig{mpsi}, \protect\fig{mpsisd} and \protect\fig{mpsidd}}
      \label{incl}
\end{figure}

Performing integration over $p_T$ for $p_T \geq 0.25\,GeV$ we find that the density of gluons $d N_G/dY  = 2 \div 2.7$ for energies $W \geq 1.8 TeV$ for $\bas = 0.2 \div 0.25$. Therefore, at high energy the dense system of partons is produced.

\subsection{Correlations and multiparton interactions (MPI)}
The value of the correlation function $R\Lb y_1,y_2\Rb$ we have discussed in the introduction (see \eq{R})).
Noticing that $\sigma_{in} = 58 mb$ for $W=7 TeV$ in our approach, we obtain that $R\Lb y_1,y_2\Rb\,\approx \,3$ for $\sigma_{eff} = 15 mb$. 
These estimates depend on the value of $N_0$ in \eq{SA}. For example, changing $N_0 $ from $N_0=0.5$ for which all previous estimates were done, to $N_0 = 0.1$ does not change the density of gluons but the value of $\sigma_{in}$ becomes $\sigma_{in} = 37 mb$ leading to $R\Lb y_1,y_2\Rb\,\approx \,1.5 \div 2$.

Hence we can claim that in the extreme case of dilute-dilute system scattering: dipole-dipole interaction at high energy, the dense and strongly correlated system of gluon is  produced.

In this system gluons are mostly originated from many parton showers. For example , the large ratio $\sigma_{in}/\sigma_{eff} \approx  2$ indicates that  the probability of two parton showers production is larger than one parton shower. 
We can calculate the inclusive production of $n$-pair of jet using our approach. Parameterizing the inclusive cross section of $n$-pair production\footnote{For simplicity we consider all $n$ pairs of jets being non identical.} in the spirit of \eq{XSEFF}, namely,
\beq \label{RN}
\frac{d^n\sigma}{ \prod^n_{i=1} d y_i d^2 p_{T,i}}\,\,=\,\,\frac{1}{\Lb \sigma^{(n)}_{eff}\Rb^n}\prod^n_{i=1} \frac{d\,\sigma}{d y_i \,d^2 p_{T,i}}
\eeq 
Using \eq{ADDAPR} one can see that 
\beq \label{XSN}
\frac{1}{\Lb \sigma^{(n)}_{eff}\Rb^n}\,\,=\,\,\Bigg\{ \frac{1}{n!} \Lb a + (1 - a)\,n!\Rb^2\Bigg\} \int d^2 b \,S^n\Lb b \Rb\,\,=\,2 \pi \Lb \frac{m^2}{2 \pi}\Rb^n \frac{1}{(n m )^2}\Bigg\{ \frac{1}{n!} \Lb a + (1 - a)\,n!\Rb^2\Bigg\} \eeq

Using $m = 0.86 \,GeV$ and $a =  0.65$  we estimate the values of $\sigma^{(n)}$ obtaining
$\sigma^{(2)}_{eff}\,= \,15 \,mb, \sigma^{(3)}_{eff}\,=\,9\,mb, \sigma^{(4)}_{eff}\,=\,5.7 \,mb$.
These numbers  illustrate  that the production of  large number of parton cascade gives the main contribution at high energy. Therefore, the large gluon  density in  dipole-dipole high energy scattering stems from the production of numerous parton  showers.
\section{Elastic slope}
Looking at  \fig{diff}, we can not escape the feeling that the dipole is similar to the proton, having total and diffractive cross sections qualitatively similar to the cross sections of the proton-proton interaction at high energies.   However, we face a problem with the shrinkage of the diffraction peak which was observed experimentally in proton-proton interaction  and, at  first sight, which is not expected in dipole-dipole scattering. Indeed, the BFKL Pomeron is not moving singularity (it is a standing branch point) and we do not expect the shrinkage of the diffraction peak for the single Pomeron exchange. On the other hand, the multi Pomeron exchanges and interactions induce the effective shrinkage.  These exchanges and interactions started to slow down the increase of the scattering amplitude due to BFKL Pomeron exchange  at 
\beq \label{B1}
N_\pom\Lb z, b \Rb\,\,\propto\,\,e^{ - m b_0  + \h z}\,\,\approx \,\,N_0 \,\,<\,\,1
\eeq
leading to 
\beq \label{b0}
b_0\,\,=\,\,\frac{1}{ 2\, m}\,z\,\,=\,\,\frac{2 \bas}{m}\,\,Y
\eeq
  \begin{figure}
  \begin{center}
      \includegraphics[width=9cm]{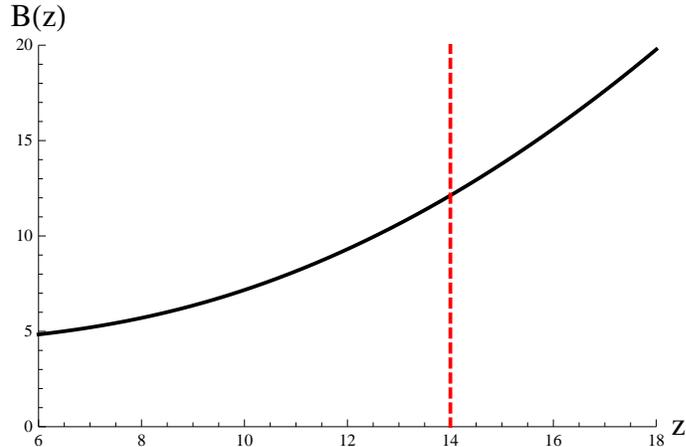}
   \end{center}
      \caption{ The elastic slope $B\Lb z\Rb$ versus $z$. $\bas$ is taken to be equal to  0.2. The vertical line shows the LHC energy range $ W = 7\,TeV$.}
      \label{b}
\end{figure}
In \fig{b} we plot the value of the elastic slope  which is equal to
\beq \label{B2}
B\Lb z \Rb\,\,\,=\,\,\, \h \frac{\int\,b^2\, d^2 b \,N_{\mbox{dipole-dipole}}}{\int\, d^2 b\, N_{\mbox{dipole-dipole}}}
\eeq
One can see that the slope is rather large and increases with the energy.  \fig{b} as well as \fig{diff} encourage us develop the description of proton-proton interaction a high energy based on CGC/saturation approach.  However, such an approach could be only phenomenological at the moment since we do not have theory of the confinement. We are going to develop such approach in the nearest future.
\section{Conclusions}
The main physical result of this paper is to demonstrate that the dense system of partons (gluons) can be produced in dilute-dilute system scattering. We illustrated this using the extreme case of dipole-dipole scattering. This increase in density is originated by   the intensive gluon cascades that can be described by the enhanced BFKL Pomeron diagrams (Pomeron loops).

For the first time    we found the analytical solution to the equation for diffraction production proposed in Ref.\cite{KOLE} using the simplified BFKL kernel.  Having this solution as well as the solution to Balitsky-Kovchegov equation we developed technique that allowed us to calculate the total cross section, cross sections for single and double diffraction in the MPSI approximation\cite{MPSI}. Hence we can discuss physics of the dilute-dilute parton system at high energy. Calculating inclusive production and two gluon correlations we see that the dense and strongly correlated system of gluons can be produced at high energy in the dipole-dipole scattering.

It should be stressed that using the BFKL Pomeron calculus and the MPSI approximation we satisfy the  $t$-channel unitarity constraint at every stage  of our calculations  and demonstrate that the resulting  scattering amplitude does not contradict  the $s$-channel unitarity. Generally speaking we have, at the moment, two approaches for the high parton density QCD: Colour Glass Condensate (CGC) approach and the BFKL Pomeron calculus. However, in the frame of the MPSI approximation these approaches give the same amplitude since, as it was shown in Ref.\cite{RISE},  the gluon cascade initiated by one dipole is the same in both.

\begin{figure}
  \begin{center}
      \includegraphics[width=8cm]{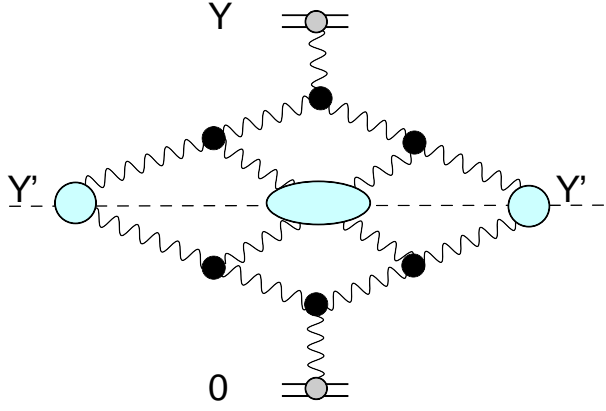}
   \end{center}
      \caption{ Corrections to the MPSI approximation.}
      \label{corr}
\end{figure}

However, at very large values of $Y$ we cannot trust the MPSI approximation. Indeed, when density of dipoles  at rapidity $Y'$  become large not only one dipole from the upper Pomeron can interact with the dipole in low Pomeron in \fig{corr} but the interaction two and more dipoles can be essential( see \fig{corr}). In terms of the BFKL Pomeron calculus it means that not only triple Pomeron interaction should be taken into account but also multiPomeron vertices have to be included.

Choosing $Y'=Y/2$ we estimate the first four Pomeron interaction (see \fig{corr}), which has not been taken into account, to demonstrate the region of validity for the MPSI approximation.
We do not know the four Pomeron vertex but it is proportional to $\bas^4$, or in other words, it has the same order of magnitude as   $\gamma^2$ (see \eq{POMTCH}). In our estimates we replace the four Pomeron vertex by $\gamma^2$ and introduce the  parameter $R\Lb z, b\Rb$ which characterizes the strength of the contribution of the four Pomeron term in the scattering amplitude and which takes the form
\beq \label{RPAR}
R\Lb z, b\Rb\,\,=\,\frac{1}{T^2\Lb 0,0\Rb}\,\int d^2 b' \, T^2\Lb z/2, \vec{b} - \vec{b}' \Rb\,T^2\Lb z/2,  \vec{b}'\Rb\,S^2\Lb z, b\Rb\Bigg)\,\,=\,\,R_1\Lb z, b\Rb\,R_2\Lb z, b \Rb
\eeq
The factor in front is need for a  correct normalization  for four Pomeron vertex being equal to $\gamma^2$.

The survival probability $S^2$ is equal to
\beq \label{S2}
S^2\Lb z, b\Rb\,\,=\,\,-\h \frac{d^2 \,N_{\mbox{dipole-dipole}}^{\mbox{appr}}\Lb T\Rb}{d T^2}{\Bigg |}_{T = T\Lb z, b \Rb}
\eeq

\begin{figure}
  \begin{tabular}{c c  c}
      \includegraphics[width=6cm]{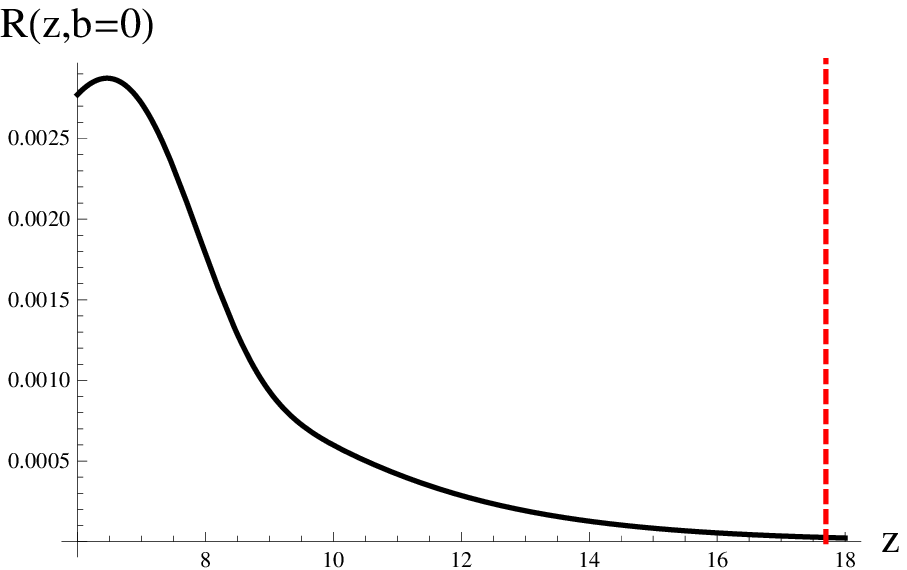}& \includegraphics[width=6cm]{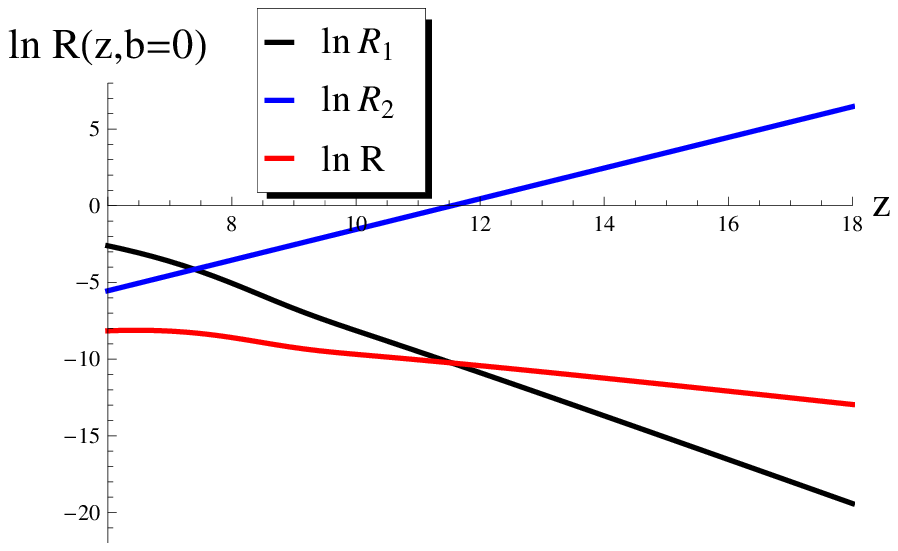}& \includegraphics[width=6cm]{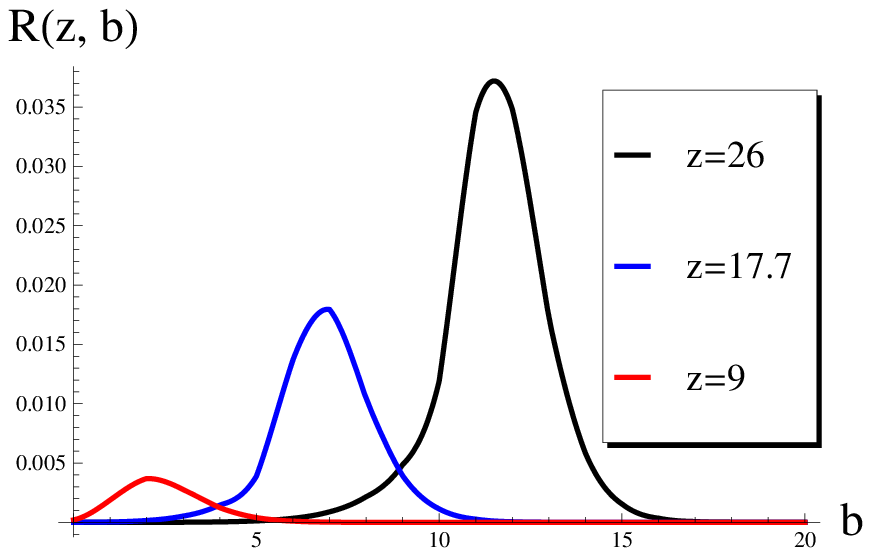}  \\
      \fig{r} - a &\fig{r}-b~& \fig{r} - c\\
       \end{tabular}
      \caption{ $R\Lb z, b = \Rb$ versus $z$ ( see  \protect\fig{r}-a and  \protect\fig{r}-b) at $b=0$  and versus $b$ at different $z$ (see  \protect\fig{r}-c) . $\bas$ is taken to be equal to  0.25. The vertical line shows the LHC energy range $ W = 7\,TeV$.}
      \label{r}
\end{figure}

This parameter is plotted in \fig{r}.  One can see that it falls down at large $z$. It happens because $R_1 \,\propto\,\,T^2$ while $S^2\,\propto\,\,1/T^3$ at large $T$.

\fig{r}-c  illustrates the $b$ dependence of $R\Lb z,b\Rb$. One can see that $R$ has a maximum at fixed $b$ but in spite of this maximum the accuracy of our calculation us less than $4\%$ even at  large values of $z(Y)$. Although it should be noted, the maximum of $R\Lb z,b\Rb$ increases with the energy and it could reach the value of several percents. Since the maximum of $R$ increases with $Y$ we see that at high energies the corrections to the MSPI approximation become essential but our estimates show that it would happen for higher energies than the LHC one.  We can safely use the MPSI approximation for the entire region of accessible energies (W $\leq$\,57 \,TeV).

Bearing in mind all the assumptions that have been made: simplified BFKL kernel, MPSI approximation, phenomenological input for impact parameter dependence, we consider this paper as the first try to approach dilute-dilute scattering theoretically.

Using the simplified BFKL kernel we were able to introduce the non-perturbative corrections at large impact parameters and the only phenomenological parameter which describes the large $b$ behaviour of the scattering amplitude, we extracted from the experimental data on double parton interaction as we discussed in the introduction.

  
  \section{Acknowledgements}
  

     We thank our    colleagues at UTFSM and Tel Aviv university for encouraging discussions.  Our special thanks go to T. Altinoluk, C. Contreras, A. Kovner, M. Lublinsky and A. Shulkin for hot and elucidating  discussions of the interrelation between BFKL Pomeron calculus and CGC approach.

     This research was supported by the  Fondecyt (Chile) grant 1100648.

 \end{document}